%% file: speckle_pose.tex
\documentclass[sigconf, anonymous=false]{acmart}
\AtBeginDocument{%
  }

\setcopyright{acmcopyright}
\copyrightyear{2018}
\acmYear{2018}
\acmDOI{XXXXXXX.XXXXXXX}

\acmConference[Conference acronym 'XX]{Make sure to enter the correct
  conference title from your rights confirmation email}{June 03--05,
  2018}{Woodstock, NY}
\acmISBN{978-1-4503-XXXX-X/18/06}

\input{packages.tex}

\begin{document}
\input{abbreviations.tex}

\title{SpecTrack: Learned Multi-Rotation Tracking via Speckle Imaging} 

\author{Ziyang Chen}
\affiliation{%
  \institution{University College London}
  \city{London}
  \country{United Kingdom}
}
\email{ziyang.chen.22@ucl.ac.uk}

\author{Mustafa Doğa Doğan}
\affiliation{%
  \institution{Adobe Research}
  \city{Basel}
  \country{Switzerland}
}
\email{doga@adobe.com}

\author{Josef Spjut}
\affiliation{%
  \institution{NVIDIA}
  \city{Durham}
  \state{NC}
  \country{USA}
}
\email{jspjut@nvidia.com}

\author{Kaan Akşit}
\affiliation{%
  \institution{University College London}
  \city{London}
  \country{United Kingdom}
}
\email{k.aksit@ucl.ac.uk}

\settopmatter{authorsperrow=4}
\renewcommand{\shortauthors}{Chen et al.}

\begin{abstract}

Precision pose detection is increasingly demanded in fields such as personal fabrication, \VR, and robotics due to its critical role in ensuring accurate positioning information.
However, conventional vision-based systems used in these systems often struggle with achieving high precision and accuracy, particularly when dealing with complex environments or fast-moving objects.
To address these limitations, we investigate \LSI, an emerging optical tracking method that offers promising potential for improving pose estimation accuracy.
Specifically, our proposed \SpecTrack leverages the captures from a lensless camera and a retro-reflector marker with a coded aperture to achieve multi-axis rotational pose estimation with high precision.
Our extensive trials using our in-house built testbed have shown that \SpecTrack achieves an accuracy of $\mathbf{0.31^\circ}$ (std=$0.43^\circ$), significantly outperforming state-of-the-art approaches and improving accuracy up to $200\%$.
\end{abstract}



\settopmatter{printacmref=false}
\setcopyright{none}
\renewcommand\footnotetextcopyrightpermission[1]{}
\pagestyle{plain}
\maketitle

\section{Introduction}
Speckle patterns appear as a random mixture of white and black dots when coherent laser light is reflected off optically rough surfaces and the interference is captured by an imaging sensor.
\LSI is used in \HCI applications such as hand gesture recognition and material identification~\cite{zizka2011specklesense, dogan2021sensicut}.
Recently, \LSI's rotation angle retrieval has been investigated through analytical methods under controlled environments~\cite{gibson2024towards,heikkinen2024self}. 
\textit{The outstanding research problem for \LSI's rotational tracking abilities is the inability to reliably detect \textbf{multi-axis absolute rotations} of objects with \textbf{minimum hardware comlexity} in a dynamic setting.}
\begin{figure}
  \includegraphics{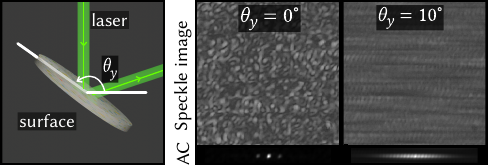}
  \caption{(left) The y-axis rotation demonstration. (right) Top: Photographs showing the speckle patterns when the target object is rotated around the y-axis at different angles. Bottom: The autocorrelation (AC) results of the photographs in the top row (enlarged for visualization).}
    \label{fig:rotation_speckle_ac}
\end{figure}
We present \SpecTrack, a rotational pose tracking system that uses \LSI to capture absolute rotations in multiple axes.
Firstly, we implement a testbed \LSI system using a lensless camera, a multi-wavelength laser, and a retroreflective marker with a coded aperture to track an object.
Inline with the findings of previous studies, the test data indicate that the multi-wavelength laser generates multiple identical speckle patterns that overlap, as shown in Fig.~\ref{fig:rotation_speckle_ac}.
Leveraging these speckle patterns, we developed a learned method that estimates the absolute rotations across multiple axes from the speckle images captured by our lensless camera.
Specifically, we make the following contributions:
\begin{itemize}[leftmargin=*]
  \item Speckle Imaging Testbed. 
  We introduce a specialized automated testbed Fig.~\ref{fig:testbed} to capture speckle images of a retroreflective surface using a lensless camera and a multi-wavelength laser.
  The testbed is capable of sub-degree motion of the target surface in two axes of rotation and one axis of translation.  
  By orienting our retroreflective surface at different inclinations, we collected a dataset that is used to train and validate our method.
  \item Learned Rotation Model.
  Based on multi-wavelength speckle overlap observations, we used a shallow network that identifies the absolute rotations of the target surface in motion from our captured speckle images, achieving a frame rate of \textbf{30 \FPS}. 
  We evaluate the performance of the rotational pose tracking system under various distances ($16~cm$ to $28~cm$) and rotation speeds (1-10 \rpm) using our testbed.
  Our multi-axis learned rotation model offers up to \textbf{$\times 2$ improvements in estimation accuracy} over the single-axis state-of-the-art~\cite{gibson2024towards}.
\end{itemize}
Our code and dataset are at \href{https://github.com/complight/SpecTrack}{\textbf{GitHub:complight/SpecTrack}}.
 
\section{Method}
\begin{wrapfigure}{l}{0.17\textwidth}
  \vspace{-0.3 cm}
  \includegraphics{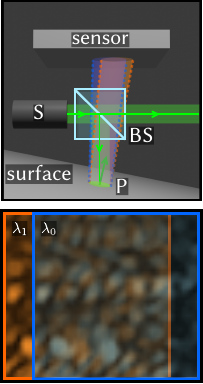}
  \caption{Schematics of the light paths (top).
  Demonstration of the overlapping speckles (bottom).
 }
 \vspace{-0.5 cm}

  \label{fig:wavelength+lightpath}
\end{wrapfigure}

We aim to remotely obtain multiple absolute rotation angles from a coded retroreflective marker by utilizing the overlapping patterns generated by the multi-wavelength laser.
\paragraph*{Overlapping Speckle Formation.} As shown in Fig.~\ref{fig:wavelength+lightpath},  
the laser beam from the source ($S$) hits an arbitrary point ($P$) and diffracts at slightly different angles due to the different wavelengths ($\lambda_0$ and $\lambda_1$).
This phenomenon shows a correlation between the surface rotation angle and the captured speckle image.

\vspace{0.5cm}

\begin{wrapfigure}{l}{0.13\textwidth}
  \includegraphics{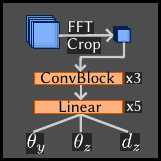}
  \caption{\NN.
 }
 \vspace{-0.5 cm}

  \label{fig:nn}
\end{wrapfigure}

\paragraph*{Sensor structure.}
The top figure in Fig.~\ref{fig:wavelength+lightpath} shows the structure of the proposed sensor, which contains a bare sensor, laser source and beam splitter ($10~mm \times 10~mm$).
The beam splitter is placed in front of the bare imaging sensor to ensure that most of the light reflected from the marker covers a large area of the sensor.
Additionally, this co-axial optical layout eliminates the light source's lateral offsets, simplifying the speckle behavior in the rotations.
We enable the absolute z-axis rotation tracking by placing a coded aperture on the retro-reflector marker (see supplementary Fig.~1).
This mask creates stripe patterns in the frequency domain (see supplementary Fig.~2) that rotate synchronously with the marker, which helps the model estimate absolute z-axis rotations.
\paragraph*{Dataset Collection.}
Since capturing samples in all six \DOF simultaneously is physically difficult, we focus on capturing the speckle imaging as the marker rotates in the z-axis and y-axis.
We add controlled close-loop motors to a rotary stage to automatically capture the speckle images when the marker is rotated in various axes, as shown in Fig.~\ref{fig:testbed}.
During the data collection, we control the motors to rotate the marker from $0^\circ$ to $40^\circ$ on the y-axis and $0^\circ$ to $90^\circ$ the z-axis.
Besides the rotations, we repeat the experiment in different depths from $16~cm$ to $28~cm$.
\paragraph*{Training Process.}
We employ a shallow \NN to handle the non-linearities of physical aspects and estimate the absolute rotation angles from speckle patterns, as shown in Fig.~\ref{fig:nn}. 

Firstly, we preprocess the captured monochrome speckle frames $I_{speckle}$ ($640\times360$~px) by transforming them into the frequency domain $\mathcal{F}(I_{speckle})$ using the \FFT.
Then the frames are central cropped and concatenated into a tensor $[\mathcal{F}(I_{\text{speckle}, i})]_{i=1}^5$ with a shape of $(5,320,180)$.
From our practical experiences, this concatenated frame tensor provides more robust results when the marker is in motion because it incorporates temporal information.
After that, we feed the samples into three convolutional blocks, each comprising a 2D convolution layer, batch normalization, ReLU activation function, and max pooling.
After the convolution, the sample is flattened and inputted into a \MLP containing six linear layers each layer is followed by a batch normalization and ReLu activation function. 
The final layer of \MLP outputs the rotation angles $\theta_y$, $\theta_z$ and the arbitrary depth $d_z$.

\section*{ANALYSIS AND RESULTS}
\begin{wrapfigure}{l}{0.28\textwidth}
  \includegraphics{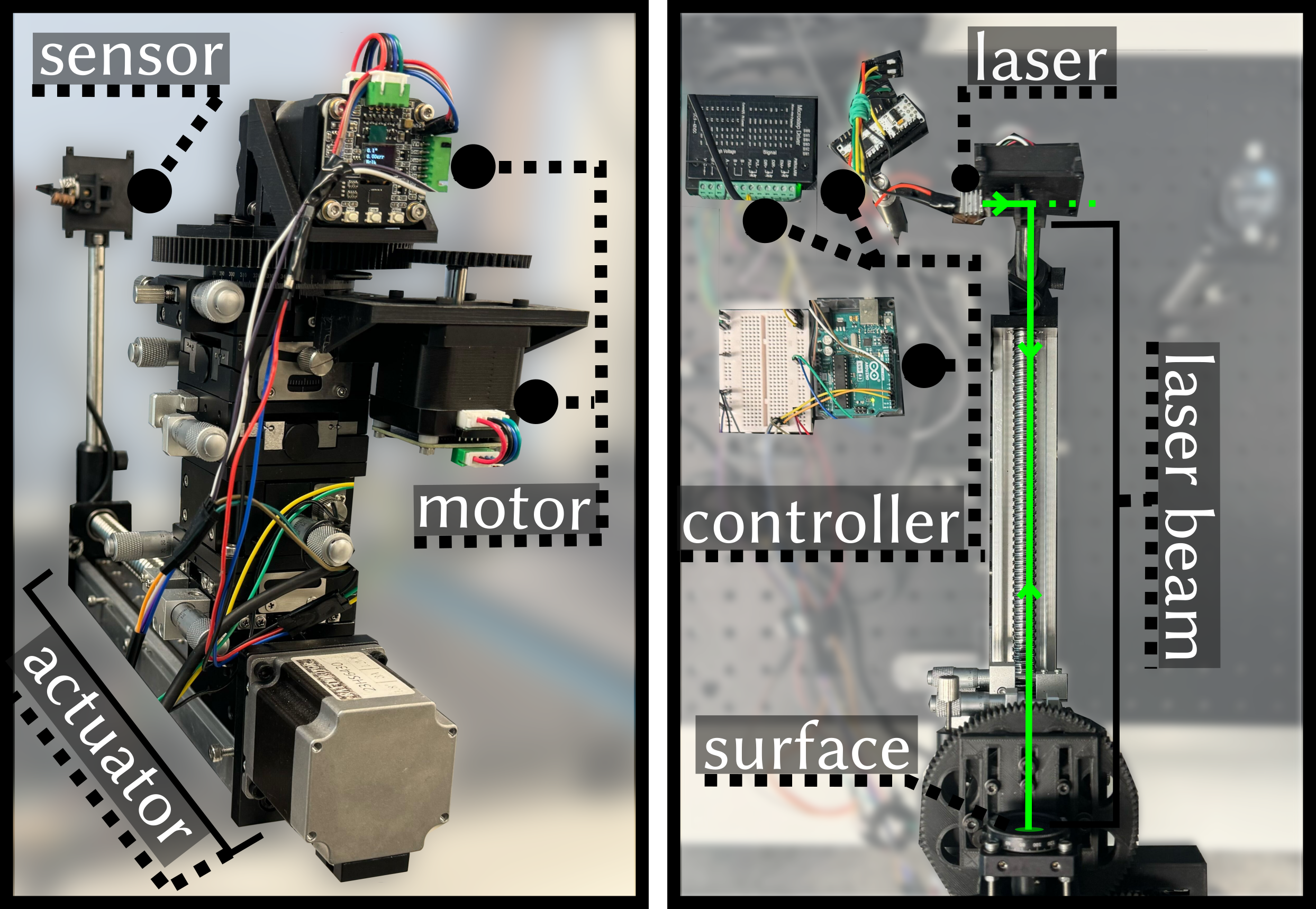}
\caption{Testbed side view (left).
Testbed top view (right).
}
\vspace{-0.5 cm}
\hspace{-5cm}
\label{fig:testbed}
\end{wrapfigure}

\begin{table}[!h]
  \vspace{-0.2cm}
  \caption{Comparison table of related works. Abbreviations: Rel.=relative, Abs.=absolute, and Acc.=accuracy.
  }
  \vspace{-0.2cm}
  \label{tab:comparison}
  \begin{tabular}{ccccccc}
    \toprule
 Method & Type & \DOF  & Sensors & Acc. ($^\circ$)\\
    \midrule
 This work                                &Abs. &\cellcolor{green!25}3   &\cellcolor{green!25}1&\cellcolor{green!25}0.3\\
\cite{gibson2024towards}   &Abs. &\cellcolor{yellow!25}2   &\cellcolor{green!25}1&\cellcolor{yellow!25}0.6\\
\cite{heikkinen2024self} &Rel. &\cellcolor{yellow!25}2   &\cellcolor{yellow!25}2     &-\\

    \bottomrule
  \end{tabular}
  \vspace{-0.5cm}
\end{table}
\paragraph*{Baseline.} 
We compare our work with the state-of-the-art from Gibson et al. \cite{gibson2024towards} under the same hardware in Fig.~\ref{fig:testbed}.
However, we lack direct access to accurate measurements, such as the wavelengths emitted by the off-the-shelf laser diode.
We subsequently employed a gradient descent-based optimization with a captured training set to get the unknown variables: dominant wavelength $\lambda_0$, wavelength differences $\Delta \lambda$, where $\Delta \lambda = \lambda_0 - \lambda_1 \ll \lambda_0$, and light source position $S$ in the 3D space.
Following this, we tested the analytical model proposed by the authors with the test set that contains the speckle images captured when the marker rotates from $0^\circ$ to $40^\circ$ in the y-axis.
The baseline result indicates the \MAE of $0.60^\circ$ ($std=0.35^\circ$) on our testbed.
\paragraph*{Results.} 
\SpecTrack achieved a lower \MAE and \std: $\mathbf{0.31^\circ}$, $\mathbf{0.44^\circ}$, respectively.
At the same time, the model can estimate the z-axis rotation with a \MAE $\mathbf{0.52^\circ}$ ($std=\mathbf{0.36^\circ}$).
Furthermore, the model adapts to varying depths, showing an accuracy of $0.15~cm$. 
\paragraph*{Future work.} 
Testing and optimizing the system in real-world environments, considering varying lighting, distances, and object motions, is crucial for successful operation in various applications including \VR, \AR, and robotics.

\bibliographystyle{ACM-Reference-Format}
\bibliography{sample-base}

\end{document}

%% file: packages.tex
\usepackage{colortbl} 
\usepackage[acronym]{glossaries}
\usepackage{xspace}
\usepackage{subcaption}
\usepackage{enumitem} 
\usepackage{wrapfig}

%% file: abbreviations.tex
\global\long\def\LSI{\gls{LSI}\xspace}
\newacronym{LSI}{LSI}{Laser Speckle Imaging}

\global\long\def\SpecTrack{\gls{SpecTrack}\xspace}
\newacronym{SpecTrack}{SpecTrack}{LSI-Based Tracking}

\global\long\def\MoCap{\gls{MoCap}\xspace}
\newacronym{MoCap}{MoCap}{Motion Capture}

\global\long\def\DOF{\gls{DOF}\xspace}
\newacronym{DOF}{DOF}{Degrees Of Freedom}

\global\long\def\NLOS{\gls{NLOS}\xspace}
\newacronym{NLOS}{NLOS}{Non Line Of Sight}

\global\long\def\IMU{\gls{IMU}\xspace}
\newacronym{IMU}{IMU}{Inertial Measurement Unit}

\global\long\def\VR{\gls{VR}\xspace}
\newacronym{VR}{VR}{Virtual Reality}

\global\long\def\AR{\gls{AR}\xspace}
\newacronym{AR}{AR}{Augmented Reality}

\global\long\def\HCI{\gls{HCI}\xspace}
\newacronym{HCI}{HCI}{Human-Computer Interaction}

\global\long\def\TOF{\gls{TOF}\xspace}
\newacronym{TOF}{TOF}{Time Of Flight}

\global\long\def\IR{\gls{IR}\xspace}
\newacronym{IR}{IR}{Infrared Radiation}

\global\long\def\SNR{\gls{SNR}\xspace}
\newacronym{SNR}{SNR}{Signal-to-Noise Ratio}

\global\long\def\MLP{\gls{MLP}\xspace}
\newacronym{MLP}{MLP}{Multi Layer Perceptron}

\global\long\def\FPS{\gls{FPS}\xspace}
\newacronym{FPS}{FPS}{Frames Per Second}

\global\long\def\FFT{\gls{FFT}\xspace}
\newacronym{FFT}{FFT}{Fast Fourier Transform}

\global\long\def\STN{\gls{STN}\xspace}
\newacronym{STN}{STN}{Spatial Transformer Networks}

\global\long\def\MAE{\gls{MAE}\xspace}
\newacronym{MAE}{MAE}{Mean Absolute Error}

\global\long\def\std{\gls{std}\xspace}
\newacronym{std}{std}{Standard Deviation}

\global\long\def\NN{\gls{NN}\xspace}
\newacronym{NN}{NN}{Neural Network}

\global\long\def\rpm{\gls{rpm}\xspace}
\newacronym{rpm}{rpm}{Revolutions per Minute}